\documentclass[11pt]{article}

\usepackage[final]{acl}
 \usepackage{microtype}
 \usepackage{booktabs}
\usepackage{makecell}
\usepackage{ulem}
\usepackage{textcomp}
\usepackage{tabularx}
\usepackage{array}
\usepackage{enumitem}
\usepackage{multirow}
\usepackage{adjustbox}
\usepackage{amsmath}
\usepackage{xcolor}
\usepackage{graphicx}
\newcolumntype{L}{>{\raggedright\arraybackslash}X}

\usepackage[english,bidi=default]{babel}
\babelprovide[import]{bengali}
\babelfont[*bengali]{rm}[Path=./]{NotoSansBengali-Regular}

\title{Exploring LLMs for South Asian Music Understanding and Generation}

\author{Faria Binte Kader \\
  University of Central Florida \\
  Florida, USA\\
  \texttt{fariabinte.kader@ucf.edu} \\\And
  Mohtasim Hadi Rafi \\
  Auburn University \\
  Alabama, USA \\
  \texttt{mzr0167@auburn.edu} \\\AND
  Shah Wasif Sazzad \\
  University of Central Florida \\
  Florida, USA \\
  \texttt{shahwasif.sazzad@ucf.edu} \\\And
  Santu Karmaker \\
  University of Central Florida \\
  Florida, USA \\
  \texttt{santu@ucf.edu} \\ }

\begin{document}
\maketitle
\begin{abstract}
Recent advancements in Large Language Models (LLMs) have shown promising results in music understanding and generation tasks. However, existing works remain confined to Western tonal traditions, offering little insight into whether current LLMs can handle structurally distinct low-resource musical traditions. We present the first systematic evaluation of LLM competence in South Asian classical music, a tradition governed by raga, tala-based melodic constraints that impose fundamentally different structural principles from Western harmony-driven music. We ground our evaluation in Hindustani classical theory and Bengali classical forms, including \textit{Rabindra} and \textit{Nazrul} Sangeet -- representative low-resource traditions within South Asian classical music. For music understanding evaluation, we introduce a \textbf{504-question-answer} benchmark spanning raga grammar, cultural knowledge, and symbolic notation reasoning, evaluating \textbf{33 LLMs} where frontier models such as \textbf{Gemini 2.5 Pro} achieve \textbf{85-90\%} accuracy, while most open-source models remain in the \textbf{23-40\%} range. For music generation, we design a five-level controlled prompting framework and find that even the strongest model produces stylistically faithful outputs only \textbf{40\%} of the time. These results reveal that structural validity and stylistic faithfulness in music generation are distinct objectives and highlight an open challenge for culturally grounded music modeling.  


\end{abstract}


\section{Introduction}

Large Language Models (LLMs) have demonstrated strong generative capabilities across a wide range of traditional NLP tasks~\cite{li2025fundamental}. Beyond NLP, they also perform well in creative and structured domains such as story generation \cite{zhao2023more}, poetry composition \cite{zhang2024llm}, code synthesis \cite{chen2021evaluating}, and mathematical reasoning \cite{lewkowycz2022solving}, showing cross-domain generalization \cite{wei2022emergent}. These capabilities have naturally extended to music. Because LLMs operate over discrete symbolic representations (i.e., tokens), symbolic music notation provides a natural interface for probing their musical capabilities. For example, ABC notation~\cite{walshaw2021abc} encodes hierarchical pitch and rhythmic structure in discrete textual form, enabling language models to process musical structure in a manner analogous to natural language. This makes symbolic notation a natural testbed for evaluating cross-domain generalization. Consequently, LLMs have been applied to both music understanding \cite{yuan2024chatmusician} and symbolic music generation tasks~\citep{ding2025songcomposer,deng2024composerxmultiagentsymbolicmusic,xing2025cocomposerllmmultiagentcollaborative}.



Despite such progress, most existing work focuses on Western musical traditions. As such, whether LLMs can understand and generate culturally specific, structurally distinct, and low-resource music (e.g., South Asian music) remains to be systematically evaluated. Indeed, existing music understanding evaluation benchmarks~\citep{yuan2024chatmusician, zhao2025abc} focus primarily on Western tonal traditions. Similarly, LLM-based music generation frameworks are designed and evaluated mostly on Western-specific properties such as harmonic chord progressions and tonal harmony~\citep{deng2024composerxmultiagentsymbolicmusic,xing2025cocomposerllmmultiagentcollaborative}. As a result, it remains unclear whether general-purpose LLMs can model musical traditions whose structure is defined by different musical principles, such as modal frameworks, cyclic rhythmic systems, and ornamentation-driven melodic movement.

Recent benchmarking efforts have increasingly moved beyond generic language understanding toward evaluating domain-specific and capability-specific competencies. For example, OmniToM evaluates whether LLMs explicitly construct belief representations rather than merely producing correct answers in Theory-of-Mind tasks, revealing limitations hidden by endpoint evaluation \cite{bawatneh2026omnitombenchmarkingtheorymind}. FinTradeBench assesses financial reasoning through the integration of heterogeneous signals such as company fundamentals and trading indicators \cite{agrawal2026fintradebenchfinancialreasoningbenchmark}.  Similarly, ITAB evaluates executable reasoning in IT automation through dynamic execution rather than static code analysis \cite{hassan2025largelanguagemodelsautomation}. These studies suggest that evaluating LLMs on specialized domains often requires task-specific benchmarks that expose reasoning abilities not captured by general-purpose evaluations. However, analogous evaluation efforts remain largely absent for culturally grounded musical traditions.

\begin{figure*}[!htb]  
    \centering
\includegraphics[width=0.9\textwidth, keepaspectratio]{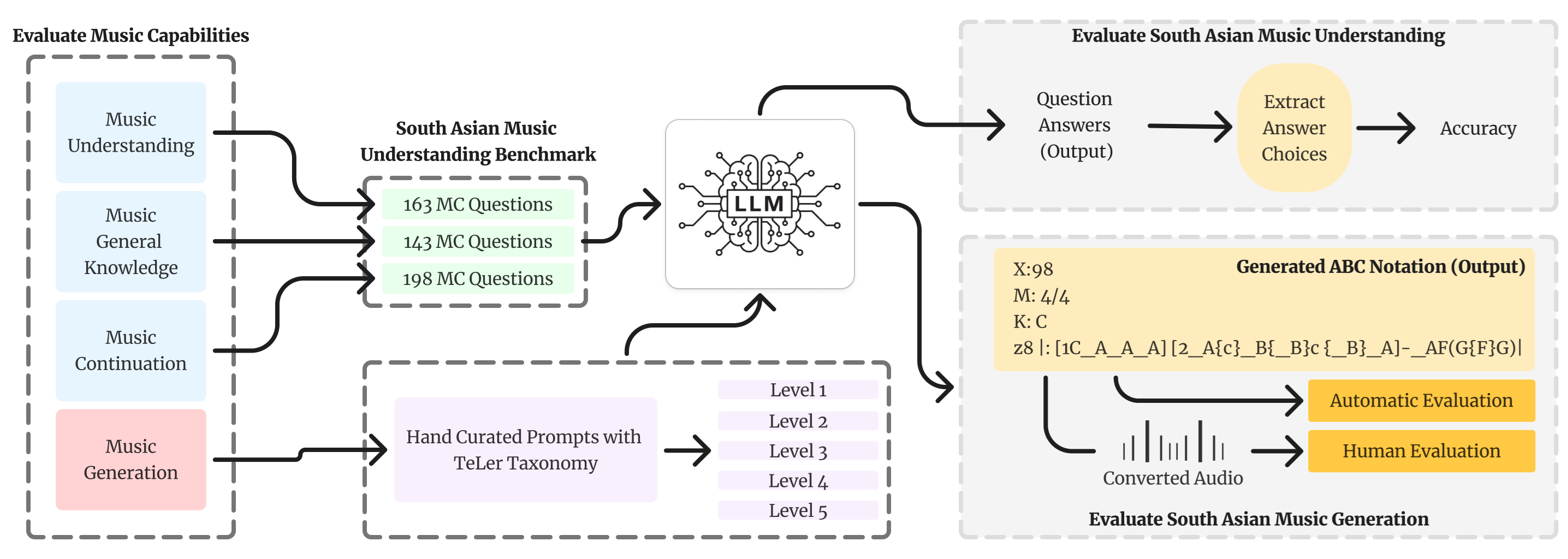}  
    \vspace{-1mm}
     \caption{LLMs' South Asian Musical Capabilities Evaluation Pipeline}
    \label{fig:pipeline}  
    \vspace{-5.5mm}
\end{figure*}

South Asian classical music is among the most well-documented non-Western traditions, making it an ideal testbed for this study. Its theoretical framework is organized around raga-based constraints \cite{pudaruth2016reflection}, cyclic tala structures \cite{rowell1986ancient}, and ornamentation-driven melodic transitions \cite{basu2023styles}, which impose long-range dependencies distinct from Western harmony-driven composition \cite{agarwal2013comparative}. Within this tradition, Bengali classical music forms such as \textit{Rabindra Sangeet} \cite{tagore_gitabitana} and \textit{Nazrul Sangeet} \cite{nazrul_institute_swarlipi} combine distinct compositional structures with raga- and tala-derived melodic rules, and lyrically driven, stylistically constrained ornamentation.

In this work, we introduce a two-part, comprehensive evaluation framework to assess LLM competence in South Asian classical music systematically (Figure \ref{fig:pipeline}). For music understanding capabilities, we construct a music understanding benchmark grounded in Bengali and Hindustani classical traditions and conduct a large-scale evaluation across \textbf{33} open-source and frontier LLMs. To evaluate the model's music generation capability, we design a \textbf{5-level} controlled prompting framework that incorporates specific Bengali classical stylistic constraints, focusing on \textit{Rabindra} and \textit{Nazrul} Sangeet, and evaluates model outputs in ABC notation using both automatic structural metrics and human judgment. Our contributions are:

\vspace{-2mm}
\begin{itemize}[leftmargin=*,itemsep=0.2ex,partopsep=0.2ex,parsep=0.2ex]

\item We introduce a \textbf{504-question benchmark} spanning three subtasks: 1) Music Theory Understanding (Hindustani classical theory as practiced in Bengali traditions including \textit{Raga Grammar}, \textit{Tala Systems}, \textit{Thaat classification}, and \textit{Ornamentation}), 2) Music General Knowledge (composers and historical figures, instrument knowledge, and regional stylistic traditions), and 3) Music Continuation (compositional reasoning in ABC notation) -- serving as the first foundational diagnostic of LLM competence in South Asian classical music in a symbolic setting. 

\item We manually curate 100 reference scores (50 Rabindra Sangeet and 50 Nazrul Sangeet) in ABC notation derived from official Swaralipi sources, serving as both continuation seeds for the benchmark and references for generation evaluation. We will publicly release the benchmark and curated reference scores.

\item We propose a 5-level prompting framework that progressively introduces musical constraints (e.g., scale, rhythm, genre, and stylistic cues) to analyze controllable symbolic music generation.

\item We evaluate 33 LLMs on the benchmark and the top 9 models on generation, revealing a large capability gap (e.g., Gemini 2.5 Pro up to 90.8\% vs. most open-source models at 23-40\% on the benchmark). Despite achieving reasonable structural validity, even the strongest model produces stylistically faithful outputs only 40\% of the time, with surface characteristics failing to capture style-specific constraints. We further show that existing automatic metrics fail to capture culturally grounded stylistic properties, highlighting the need for culturally aware evaluation of symbolic music generation.
\end{itemize}
The benchmark questions and reference songs have been made available \footnote{\url{https://github.com/Faria-Binte-Kader/South-Asian-Music-data}}.

\vspace{-1mm}
\section{Background \& Related Work}
\vspace{-1mm}
\subsection{South Asian Classical Music and Bengali Classical Traditions}
South Asian classical music follows principles that are fundamentally different from those of Western tonal systems \cite{agarwal2013comparative}. Rather than harmonic chord progressions, it is structured around raga, a melodic framework that specifies permitted pitches, expressive mood \cite{pudaruth2016reflection}, and tala. This cyclic rhythmic system organizes time through repeating beat cycles \cite{rowell1986ancient}. Melodic motion is further shaped by gamakas (raga-specific ornamentations) \cite{basu2023styles}. These principles appear across multiple Bengali musical traditions including khayal, thumri, baul, and kirtan. Among them, Rabindra Sangeet and Nazrul Sangeet are two of the most influential and well-documented forms. Rabindra Sangeet comprises 2,232 songs by Rabindranath Tagore \cite{tagore_gitabitan} and is characterized by close coupling between lyrical prosody and melodic contour, drawing from Hindustani classical, Bengali folk, and Western influences \cite{gupta2016philosophy}. Nazrul Sangeet includes nearly 4,000 compositions by Kazi Nazrul Islam \cite{nazrul_institute_swarlipi} and adheres closely to classical performance conventions \cite{v_poetry}. Their well-defined compositional rules and reliance on raga-tala structure make them a challenging testbed for evaluating culturally grounded music reasoning in LLMs.

\vspace{-2mm}
\subsection{Music representation}
\vspace{-1mm}
Music for computational processing can be represented either as transformed audio waveforms \citep{logan2000mel,takuya1999realtime} or as symbolic notations that encode musical structure \citep{rothstein1995midi, goodmusicxml, zhao2025abc, nienhuys2003lilypond}. Among symbolic formats, \textbf{ABC Notation} \cite{walshaw2021abc} encodes pitch, rhythm, meter, and key signature in compact plain text. 
ABC notation produces significantly shorter sequences than MIDI or MusicXML \cite{yuan2024chatmusician}, and LLMs perform better on ABC-encoded tasks \cite{qu2024mupt}, making it a practical choice without the complexity of audio encoding.

\vspace{-2mm}
\subsection{Music Understanding Benchmarks}
Music understanding benchmarks have evolved from narrow Music Information Retrieval (MIR) tasks such as genre classification, beat tracking, and chord estimation \cite{downie2014ten} to broader multi-task evaluations covering acoustic, semantic, and structural reasoning. Benchmarks such as MARBLE \cite{yuan2023marble} and CMI-Bench \cite{yinghao2025cmi} evaluate acoustic and semantic capabilities, while AIR-Bench \cite{yang2024air} and MuChoMusic \cite{weck2024muchomusic} extend evaluation to multimodal audio-language reasoning. For symbolic music, MusicTheoryBench \cite{yuan2024chatmusician}, ABC-Eval \cite{zhao2025abc}, ZIQI-Eval \cite{li2024musicmaestromusicallychallenged}, and WildScore \cite{mundada2025wildscore} benchmark LLM competence primarily in Western tonal traditions. However, existing benchmarks exhibit these limitations: non-Western traditions, including South Asian classical music, are largely absent; symbolic reasoning under raga-based compositional constraints is not evaluated; and culturally grounded musical structures such as raga grammar and tala organization are not represented. Although ZIQI-Eval includes limited South Asian content, it focuses only on surface-level factual questions. More broadly, recent work has questioned whether strong benchmark performance necessarily reflects the underlying capabilities that benchmarks are intended to measure. ALIGN-SIM \cite{mahajan-etal-2024-align} introduces alignment-based evaluation criteria for sentence representations and shows that models can achieve strong downstream performance despite exhibiting representational inconsistencies. Similarly, OmniToM \cite{bawatneh2026omnitombenchmarkingtheorymind} demonstrates that correct answers on Theory-of-Mind tasks may conceal failures in underlying belief modeling. These findings motivate evaluation frameworks that move beyond surface task success and examine whether models possess the underlying competencies required for robust performance. In the context of music, this raises an analogous question: whether structural correctness in symbolic music tasks reflects genuine musical understanding and stylistic competence. These gaps motivate the benchmark introduced in this work.

\subsection{Music Generation Frameworks}
Based on representations, we observe music generation and evaluation frameworks in both the audio and symbolic domains.

\paragraph{Audio Models}

Transformer-based \citep{agostinelli2023musiclm,kreuk2022audiogen} and diffusion-based audio music generation architectures \citep{yuan2025yue,liu2025songgen,ning2025diffrhythm} have advanced rapidly but remain constrained by distributional bias toward Western-centric training corpora, with performance degrading across underrepresented cultural traditions \citep{martak2026sound, 11336726, sawaengsawangarom2025deep}. Models further struggle with fine-grained controllability, and consistent text–audio alignment, while diffusion-based approaches remain computationally expensive \citep{zhao2025ai, kong2025deep, zhang2024instruct}. Objective metrics further correlate poorly with human musical judgment, and standardized cross-cultural benchmarks remain limited \citep{lemercier2025diffusion, dong2025survey, chiu2025artificial}, motivating complementary investigation in symbolic domains.

\paragraph{Symbolic Models}
In symbolic music generation, models producing scores in formats such as MIDI or ABC notation have progressed from GAN-based systems (MuseGAN \cite{dong2018musegan}) to transformer (Museformer \cite{yu2022museformer}) and diffusion-based approaches \cite{zhang2023sdmuse}, enabling multi-instrument composition. However, symbolic datasets remain small and culturally narrow \cite{wu2022exploring}, exhibit weak harmonic control \cite{li2025erld}, and lack unified objective and perceptual evaluation framework \cite{liang2025muspikebenchmarkevaluationframework}.

\paragraph{LLMs for Symbolic Music Generation and Non-Western Domains}
Text-based LLMs have extended symbolic generation capabilities, with ChatMusician \cite{yuan2024chatmusician} treating ABC notation as a second language, SongComposer \cite{ding2025songcomposer} enabling lyrics and melody generation, and ComposerX \cite{deng2024composerxmultiagentsymbolicmusic} and CoComposer \cite{xing2025cocomposerllmmultiagentcollaborative} extending this through multi-agent prompting frameworks. 
Cross-lingual transfer studies further show that English-centric pretraining underperforms on low-resource languages \cite{mehta-etal-2025-music}, a pattern that extends to the music domain, motivating us to conduct a systematic evaluation of LLM competence in South Asian symbolic musical capabilities grounded in Hindustani and Bengali classical traditions.


\vspace{-2mm}
\section{Methodology}
\vspace{-1mm}
\subsection{South Asian Music Understanding Benchmark Curation}
We curated a 504-question multiple-choice benchmark to evaluate LLMs’ capabilities in understanding South Asian music, spanning Hindustani classical music theory and Bengali classical traditions. The benchmark includes three subtasks reflecting distinct but complementary knowledge domains:

\vspace{-3mm}
\begin{itemize}[leftmargin=*,itemsep=0.2ex,partopsep=0.2ex,parsep=0.2ex]
\item \textbf{Music Theory Understanding (163 Questions)} evaluates LLMs’ comprehension of the theoretical grammar underlying Hindustani classical music, as practiced in Bengali traditions such as Raga grammar, tala systems, thaat classification, ornamentation, and notation.
\item \textbf{Music General Knowledge (143 questions)} evaluate factual and cultural knowledge grounded in Bengali and broader Hindustani musical heritage covering Composers, historical figures, instruments, and regional stylistic traditions.
\item Similar to \cite{li2024musicmaestromusicallychallenged}, \textbf{Music Continuation (198 questions)} task evaluates whether LLMs can apply structural musical knowledge in practice. Each question presents the beginning of a Rabindra or Nazrul Sangeet in ABC notation and asks the model to select the most structurally coherent continuation from four options. 
\end{itemize}
\vspace{-2mm}
Examples for each subtask from the benchmark are provided in Appendix~\ref{benchmark_examples}. All questions were reviewed by an expert in Bengali classical music.
\vspace{-1mm}
\subsubsection{Language and Extraction}

Theory and knowledge questions were curated exclusively in Bengali to preserve terminology. Models were prompted in Bengali and allowed to respond freely. Responses in Bengali or English were processed using language-agnostic option-letter extraction (A-D). Outputs were first scanned for explicit answer declarations; otherwise, standalone option letters were counted. Ties or missing predictions were marked incorrect. 
Full extraction rules are provided in Appendix~\ref{answer}.

\subsubsection{Distractor Construction}
A corpus of 100 manually transcribed ABC scores (50 Rabindra, 50 Nazrul) from official Swaralipi (Hindustani music notation system) served both as continuation seeds and generation references. Continuation distractors were constructed by sampling melodic segments from other songs within the annotated set, deliberately introducing structural mismatches through cross-genre substitution and rhythmic inconsistencies, ensuring that distractors are plausible surface-level continuations while being structurally incorrect with respect to raga consistency or tala alignment.
\vspace{-1mm}
\subsection{Music Generation Prompt Design}
Prompt formulation is known to substantially affect observed LLM performance. To systematically control prompt specificity, we adopt the TELeR taxonomy ~\citep{santu2023telergeneraltaxonomyllm}, which has been successfully used in recent evaluations across domains including educational content generation \cite{knipper2025instructionalgoalalignedquestiongeneration} and cognitive bias assessment \cite{knipper2025biasdetailsassessmentcognitive}. These studies demonstrate that increasing prompt detail can improve instruction adherence, output quality, and behavioral consistency. We therefore use TELeR to investigate how varying levels of musical guidance influence South Asian music generation.

For 50 Rabindra and 50 Nazrul lyrics, five prompt variants were used per model, yielding 500 prompts in total. Metadata from the original compositions was incorporated in natural-language form to enable controlled comparison with the references. Full prompts are provided in Appendix~\ref{prompt}.

\vspace{-2mm}
\section{Experimental Design}
\vspace{-1mm}
\subsection{Models}
\vspace{-1mm}
We evaluated 30 open-source LLMs across multiple families and parameter scales, along with 3 proprietary systems, on the music understanding task. All models were evaluated at temperature = 1.0, top-p = 1.0, top-k = -1, a fixed seed of 492, and a maximum generation length of 4096 tokens. For the music generation task, the top 9 models (scoring above 40\% on average in 3 benchmark subtasks) were selected for evaluation. QwQ-32B-Preview and Phi-4 were excluded despite meeting the threshold, due to high ABC syntax error rates. Table~\ref{tab:models} summarizes all models.
\begin{table}[t]
\centering
\footnotesize
\setlength{\tabcolsep}{2pt}
\renewcommand{\arraystretch}{1.0}
\resizebox{0.85\columnwidth}{!}{%
\begin{tabularx}{\columnwidth}{@{}lX@{}}
\toprule
\textbf{Family} & \textbf{Models} \\
\midrule
\multicolumn{2}{@{}l}{\textit{Open-source}} \\

Qwen 2.5 \cite{yang2025qwen2} 
& 0.5B--14B It \\

QwQ \cite{qwq-32b-preview} 
& 32B Preview \\

Falcon 3 \cite{almazrouei2023falcon} 
& 3B, 7B, 10B It \\

DeepSeek R1 \cite{guo2025deepseek} 
& Distill-Qwen (1.5B, 7B, 14B, 32B), Distill-Llama-8B \\

Gemma 2 \cite{team2024gemma} 
& 2B It \\

Gemma 3 \cite{team2025gemma} 
& 1B It \\

OLMo 2 \cite{olmo20242} 
& 7B, 13B It \\

OLMo \cite{groeneveld2024olmo} 
& 7B It \\

\makecell[l]{Llama 3.x\\\cite{grattafiori2024llama}} 
& 3.2 (1B, 3B), 3.1 (8B) It \\

Phi 3.5 \cite{abdin2024phi3technicalreporthighly} 
& Mini \\

Phi 4 \cite{abdin2024phi} 
& Phi-4, Phi-4 Mini \\

Mistral \cite{jiang2023mistral7b} 
& 7B It-v0.3 \\

Cogito-v1-preview \cite{cogitov1preview}
& Llama-3B,8B Preview \\

ChatMusician \cite{yuan2024chatmusician} 
& fine-tuned Llama2 \\

\midrule
\multicolumn{2}{@{}l}{\textit{Proprietary}} \\

GPT-3.5 \cite{openai2022chatgpt} & --- \\
GPT-4o \cite{hurst2024gpt} & --- \\
\makecell[l]{Gemini 2.5 Pro\\\cite{comanici2025gemini}} & --- \\

\bottomrule
\end{tabularx}}

\caption{Models evaluated for South Asian music understanding task. ``It'' stands for Instruct models.}
\vspace{-5mm}
\label{tab:models}
\end{table}

\vspace{-1mm}
\subsection{Evaluation metrics}
\vspace{-1mm}
\subsubsection{Music Understanding Evaluation}
We extracted the answer choices and calculated the accuracy scores between the ground truth and the generated answer.
\vspace{-1mm}
\subsubsection{Music Generation Evaluation}
\paragraph{Automatic Evaluation}
For the automatic evaluation of generated scores, we structure the evaluation framework into two sections: 1) reference-based evaluation and 2) reference-free evaluation. 

In \textit{Reference-Based Evaluation}, we compare the ABC notation score generated by the model against the original reference ABC notation score corresponding to the given song, using the widely used KL Divergence \cite{kullback1951kullback}.


In \textit{Reference Free Evaluation}, we evaluate the generations on whether the generated score adhered to the given prompt or not, with metrics like Adherence to scale, ABC syntax accuracy (ABC Parse Success Rate), etc. Other aspects of the prompt (e.g., genre, lyrics) were left for human evaluation due to the abstract nature of the task. We further computed feature quality metrics including Repetition rate \cite{yuan2024chatmusician} and Pitch Histogram Entropy \cite{wu2020jazz}. Appendix \ref{auto} provides details on each metric and its implementation.

\paragraph{Human Evaluation}
Automatic metrics cannot fully capture perceptual quality or stylistic authenticity. We therefore conducted a human evaluation with three annotators who had formal training in music theory and familiarity with both Rabindra and Nazrul Sangeet. A total of 180 samples were evaluated, comprising 20 randomly selected generations (with balanced representation of Rabindra and Nazrul Sangeet) per model from the top nine performing models, all generated using Level 3 prompts, ensuring enough parseable audio samples were available for each model. Level 3 was chosen because it provides explicit structural constraints without extensive stylistic scaffolding; a quantitative justification is provided in Section~\ref{level3}. Annotators followed standardized annotation guidelines and rated each sample on six criteria: Structureness, Genre analysis, Style analysis, Emotion, Adherence to the instruction, and Harmoniousness. The annotation guideline is provided in Appendix~\ref{criteria}.

Ordinal criteria (e.g., Structureness) were mapped to numeric scales (1-5) for calculation. For Genre and Style Analysis, we compared annotator-assigned labels against the prompt specification and reported accuracy as the percentage of matches. Labels were determined by majority voting across three annotators; samples with no majority agreement were counted as mismatches.

\vspace{-2mm}
\section{Results}
\vspace{-1mm}
In this section, we focus on answering several research questions: \textbf{RQ-1:} Do current LLMs understand grammatical concepts of Indian and Bengali classical music? \textbf{RQ-2:} How well do current LLMs exhibit factual and cultural knowledge of Bengali and Hindustani musical heritage? \textbf{RQ-3:} Can current LLMs generate stylistically faithful symbolic music scores adhering to the compositional conventions of Bengali classical genres? \textbf{RQ-4:} Are the current symbolic music generation evaluation metrics reliable for capturing the structural and stylistic properties of South Asian symbolic music?

\subsection{South Asian Music Understanding}
\begin{figure*}[t]
\centering
\includegraphics[width=\textwidth, keepaspectratio]{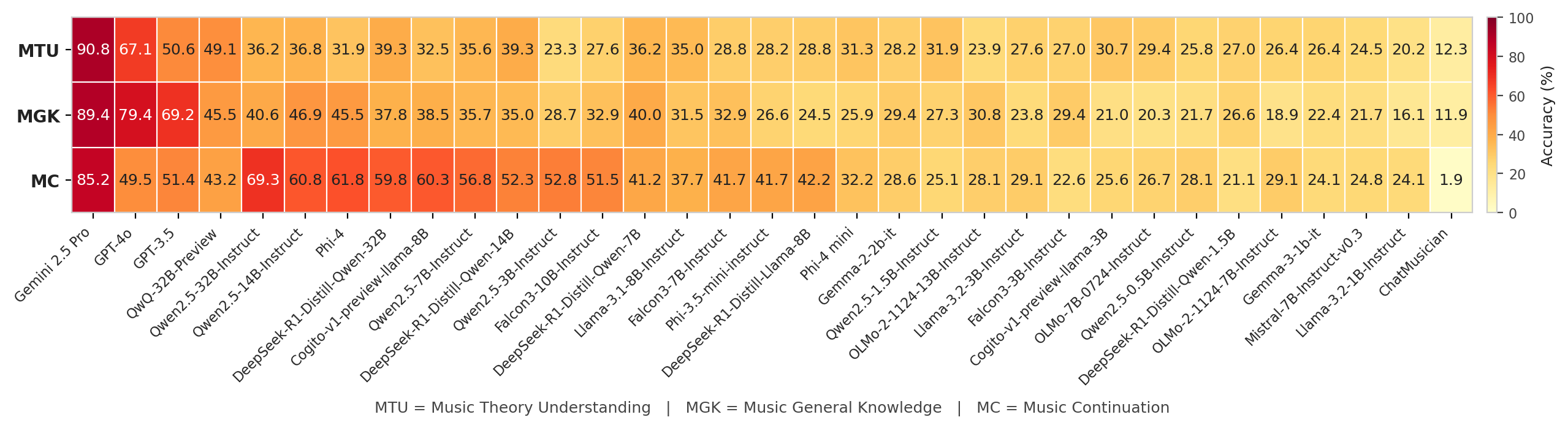}
\vspace{-7mm}
\caption{Heatmap of 33 model accuracy scores across three South Asian music understanding subtasks. Rows correspond to tasks and columns correspond to models. Darker colors indicate higher accuracy. Models are sorted by theory understanding accuracy. }
\vspace{-4mm}
\label{fig:heatmap_metrics}
\end{figure*}
Figure~\ref{fig:heatmap_metrics} presents accuracy scores across all three benchmark subtasks for the 33 evaluated models. Addressing \textbf{RQ-1} and \textbf{RQ-2}, results indicate that most open-source models show limited understanding of both the theoretical grammar and cultural knowledge underlying Hindustani and Bengali classical music, with Music Theory Understanding scores remaining in the 23-40\% range. Poor performance on Music Continuation further extends this deficit beyond theory to the practical application of South Asian musical grammar. Gemini 2.5 Pro substantially outperforms all others across all three subtasks (90.8\%, 89.4\%, 85.2\%), followed by GPT-4o and GPT-3.5, indicating that while frontier models exhibit emergent comprehension of South Asian musical knowledge, most open-source LLMs remain ill-equipped for this culturally grounded domain. Several noteworthy observations:
\begin{itemize}[leftmargin=*,itemsep=0.2ex,partopsep=0.2ex,parsep=0.2ex]
\vspace{-2mm}
    \item \textit{Scaling yields diminishing returns within families.}
    Within the Qwen2.5 family, the 32B, 14B, and 7B variants score nearly identically on Music Theory Understanding (36.2\%, 36.8\%, 35.6\%), and DeepSeek-R1-Distill Qwen 14B and 32B are identical (39.3\%), suggesting a performance ceiling regardless of scale.

    \item \textit{Music Continuation is anomalously high for some models.}
    Several mid-tier models score surprisingly high on Music Continuation despite having a poor Music Theory Understanding score (Qwen2.5-32B:
    69.3\%, Qwen2.5-14B: 60.8\%), suggesting the task may be partially solvable through surface-level pattern matching rather than genuine musical reasoning.

    \item \textit{Reasoning ability does not substitute for musical knowledge.}
    QwQ-32B-Preview, a reasoning-focused model, underperforms GPT-3.5 on Theory Understanding (49.1\% vs.\ 50.6\%), indicating that general chain-of-thought reasoning does not compensate for the absence of culturally grounded musical
    knowledge.

    \item \textit{ChatMusician completely fails on South Asian content.}
    The only music-specific fine-tuned model scores lowest across all subtasks (12.3\%, 11.9\%, 1.9\%). Output inspection reveals question repetition in Music Continuation and hallucinated Bengali non-answers, confirming that fine-tuning on Western ABC notation does not transfer to South Asian traditions.

\end{itemize}
\vspace{-2mm}

\vspace{-1mm}
\subsection{South Asian Music Generation}
\vspace{-1mm}
\subsubsection{Automatic Evaluation} \label{level3}
\begin{figure*}[t]
\centering
\includegraphics[width=\textwidth]{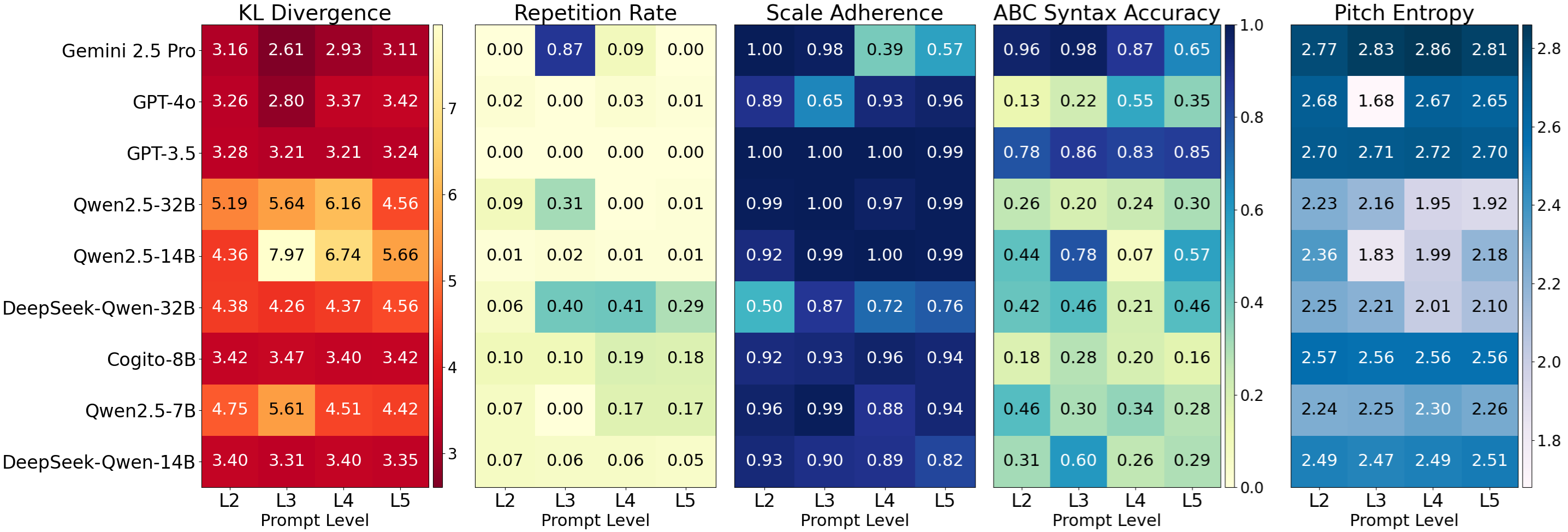}
\vspace{-7mm}
\caption{Heatmaps of automatic evaluation metrics across models and prompt levels. Rows correspond to models and columns correspond to prompt levels (L2-L5). Darker colors indicate stronger performance for each metric.}
\label{fig:heatmap_auto}
\vspace{-2mm}
\end{figure*}

Figure~\ref{fig:heatmap_auto} shows automatic evaluation scores
for the top 9 models across prompt levels L2-L5 (Level 1 excluded due to noisy, unconstrained generations; see Appendix~\ref{prompt}). Gemini 2.5 Pro achieves the best scores across all metrics, and its pitch distributions most closely match the reference entropy of \textbf{2.86}. Among open-source models, Cogito-v1-preview-llama-8B
and DeepSeek-R1-Distill-Qwen-14B show relatively competitive KL Divergence scores, though both are considerably lagging behind Gemini 2.5 Pro.

Levels 2 and 3 consistently achieve competitive scores across all metrics (Table~\ref{tab:mean_auto_eval} shows mean scores across levels), while Levels 4
and 5 yield no appreciable improvement despite additional contextual detail, suggesting that beyond a moderate level of prompt specificity, additional information does not meaningfully enhance generation quality. Level 3 was therefore selected as the representative prompt level for human evaluation.

\vspace{-1mm}
\subsubsection{Human Evaluation}
\vspace{-1mm}
Table~\ref{tab:human_eval} presents human evaluation results.
Gemini 2.5 Pro achieves the highest scores across all ordinal
criteria (3.25-3.65), followed by Cogito-v1-preview-llama-8B.
Most open-source models score below 2.5, indicating that
structurally parseable outputs do not necessarily translate to
perceptually coherent or emotionally aligned melodies.

\begin{figure}[!htb]
\vspace{-1mm}
    \centering
\includegraphics[width=0.9\columnwidth]{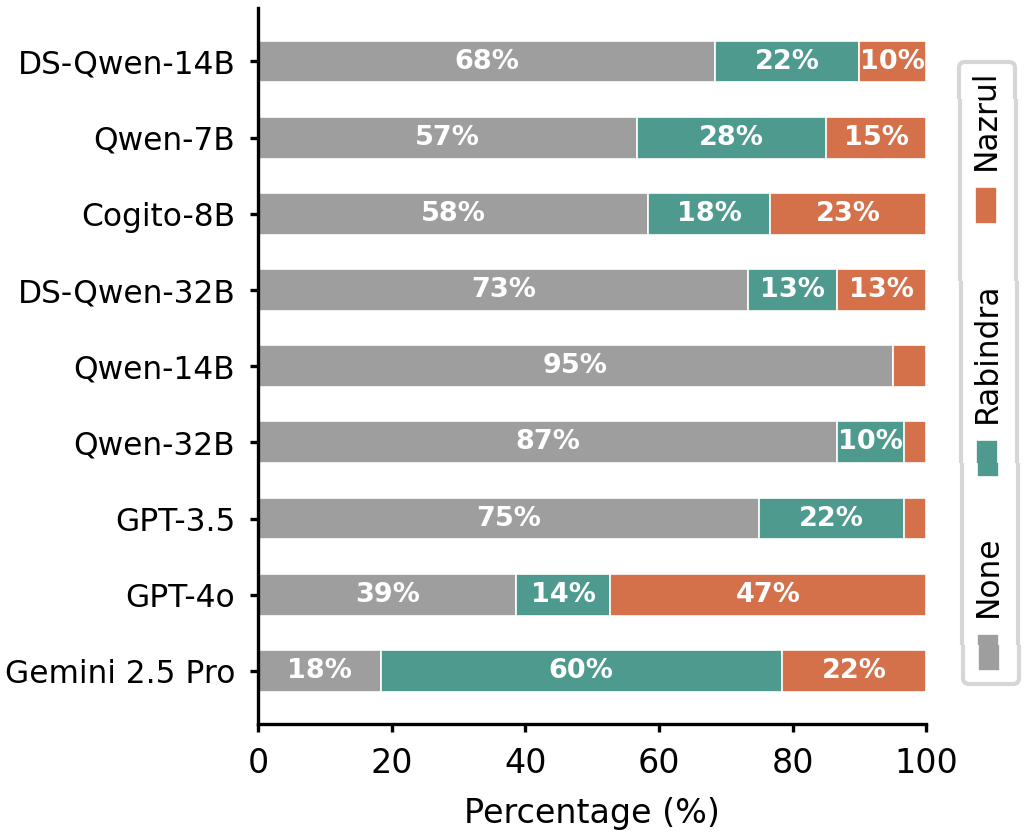}  
    \vspace{-2mm}
     \caption{Percentage of generated Styles (Rabindra, Nazrul, None) for each. Here, DS-Qwen = DeepSeek-R1-Distill-Qwen, Qwen = Qwen 2.5 instruct models}
    \label{fig:stylepercentage}
  \vspace{-4mm}  
\end{figure}

For Genre and Style Accuracy, Gemini 2.5 Pro maintains strong
adherence to the broader South Asian classical genre (95\%) but
correctly generates the Rabindra or Nazrul Sangeet style in only
40\% of cases, revealing the central finding for \textbf{RQ-3}.
Figure~\ref{fig:stylepercentage} shows that for most models,
the majority of generated outputs fall into no recognizable category, suggesting that current LLMs can approximate the surface characteristics of South Asian classical music but lack the compositional specificity needed for faithful stylistic control. It is worth noting that ABC notation cannot approximate some raga-specific gamakas such as meend and microtonal inflections, which may partly contribute to the gap between structural validity and stylistic faithfulness beyond model limitations.

\begin{table*}[!htb]
\centering
\footnotesize
\resizebox{0.85\textwidth}{!}{%
\begin{tabular}{lcccccc}
\toprule
\textbf{Model} &
\makecell{\textbf{Structureness}\\\textbf{(1-5)} $\uparrow$} &
\makecell{\textbf{Emotion}\\\textbf{(1-5)} $\uparrow$} &
\makecell{\textbf{Adherence}\\\textbf{to Instruction (1-5)} $\uparrow$} &
\makecell{\textbf{Harmoniousness}\\\textbf{(1-5)} $\uparrow$} &
\makecell{\textbf{Genre}\\\textbf{Accuracy} $\uparrow$} &
\makecell{\textbf{Style}\\\textbf{Accuracy} $\uparrow$} \\
\midrule
Gemini 2.5 Pro               & \textbf{3.65} & \textbf{3.42} & \textbf{3.25} & \textbf{3.53} & \textbf{0.95} & \textbf{0.40} \\
GPT-4o                       & 2.96 & 2.18 & 2.35 & 2.77 & 0.75 & 0.40 \\
GPT-3.5                      & 2.43 & 2.12 & 2.22 & 2.97 & 0.50 & 0.25 \\
Qwen2.5-32B-Instruct         & 1.50 & 1.33 & 1.55 & 1.62 & 0.05 & 0.05 \\
Qwen2.5-14B-Instruct         & 1.95 & 1.35 & 1.60 & 2.05 & 0.00 & 0.00 \\
DeepSeek-R1-Distill-Qwen-32B & 2.07 & 2.18 & 2.17 & 2.25 & 0.55 & 0.25 \\
Cogito-v1-preview-llama-8B   & 3.30 & 3.00 & 3.08 & 3.35 & 0.65 & 0.30 \\
Qwen2.5-7B-Instruct          & 2.33 & 2.53 & 2.25 & 2.32 & 0.55 & 0.40 \\
DeepSeek-R1-Distill-Qwen-14B & 2.07 & 2.33 & 2.05 & 2.30 & 0.60 & 0.25 \\
\bottomrule
\end{tabular}}

\caption{Human evaluation results on 6 criteria. Columns 2-5 show average annotator scores on ordinal criteria (1-5), columns 6-7 show majority-vote accuracy for Genre and Style Analysis.}
\label{tab:human_eval}
\vspace{-3mm}
\end{table*}

\begin{figure*}[!htb]  
    \centering
\includegraphics[width=0.9\textwidth, keepaspectratio]{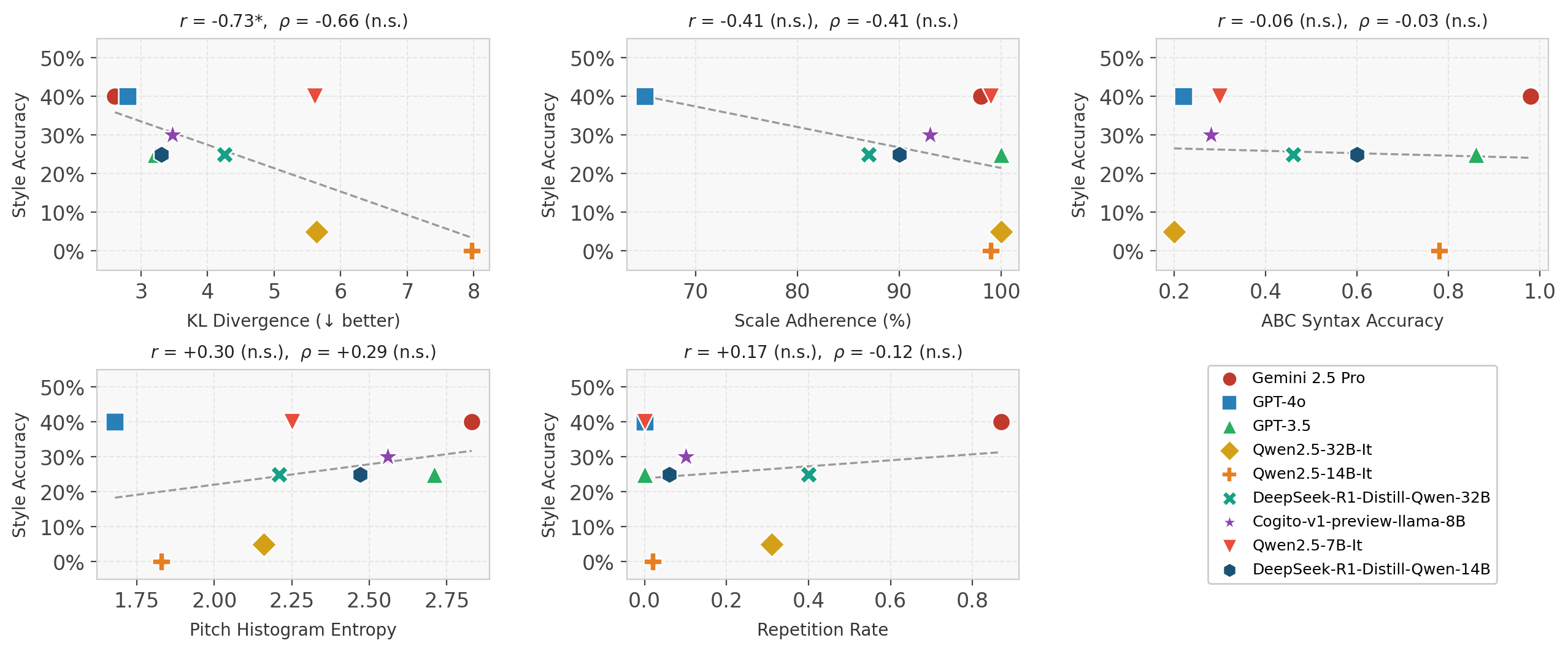}  
    \vspace{-2mm}
     \caption{Pearson and Spearman correlations between each automatic evaluation metric and human Style Accuracy across nine models (Level 3 generations). Each point represents one model; the dashed line shows the linear trend. * indicates $p< 0.05$ and n.s. indicates not significant.}
    \label{fig:style}
  \vspace{-5mm}  
\end{figure*}

\begin{table}[!htb]
\centering
\footnotesize
\setlength{\tabcolsep}{4pt}
\resizebox{0.85\columnwidth}{!}{%
\begin{tabular}{lccc}
\toprule
\textbf{Criteria} & \textbf{$\alpha$} & \textbf{$\kappa$} & \textbf{Majority agreement} \\
\midrule
Structureness       & 0.437 & 0.441 & 83.2\% \\
Emotion             & 0.325 & 0.347 & 78.8\% \\
Adherence           & 0.268 & 0.282 & 77.1\% \\
Harmoniousness      & 0.430 & 0.431 & 81.6\% \\
\midrule
Genre Analysis      & 0.152 & 0.151 & 100\% \\
Style Analysis      & 0.206 & 0.204 & 92.2\% \\
\bottomrule
\end{tabular}}

\caption{IAA across 6 criteria. $\alpha$: Krippendorff's alpha, $\kappa$: mean weighted Cohen's Kappa (ordinal) / Fleiss' Kappa (nominal).}
\vspace{-7mm}
\label{tab:iaa_results}
\end{table}
We computed Inter-Annotator Agreement (IAA) across 3 annotators using 
Krippendorff's $\alpha$ \cite{krippendorff2011computing}, mean 
weighted Cohen's $\kappa$ \cite{cohen1960coefficient} for ordinal criteria (Structureness, 
Emotion, Adherence, Harmoniousness, rated on a 1-5 scale), and 
Fleiss' $\kappa$ \cite{fleiss1971measuring} for nominal criteria (Genre Analysis and Style 
Analysis, treated as unordered categorical labels). Majority 
agreement is reported as the percentage of items where at least 
2 of 3 annotators assigned the same label. For ordinal criteria, both $\alpha$ 
and $\kappa$ indicates moderate agreement for Structureness 
($\alpha=0.437$, $\kappa=0.441$) and Harmoniousness 
($\alpha=0.430$, $\kappa=0.431$), suggesting annotators share 
reasonable common understanding of structural and harmonic quality, with majority agreement of 83.2\% and 81.6\% respectively. Adherence to instruction and Emotion show weaker agreement 
($\alpha=0.268$-$0.325$), reflecting the inherently subjective 
nature of these criteria. For nominal criteria, Genre Analysis 
and Style Analysis show low $\alpha$ and $\kappa$ 
($\alpha=0.152$, $\kappa=0.151$; $\alpha=0.206$, $\kappa=0.204$), Yet the majority agreement remains high at 100\% and 92.2\% respectively. This divergence indicates that the low $\alpha$ is driven by the highly skewed label distribution of the generated outputs rather than by genuine annotator disagreement, since annotators consistently converged on the same label. Still, that label was predominantly one category, confirming that models genuinely failed to generate stylistically distinctive outputs rather than annotators being unable to classify them. Thus answering \textbf{RQ-3}: current LLMs demonstrate partial competence in structural generation but fall substantially short of faithful stylistic control in Bengali classical genres, posing a critical open challenge for culturally grounded music modeling.

\begin{figure}[!htb] 
\vspace{-2mm}
    \centering
\includegraphics[width=0.44\textwidth, keepaspectratio]{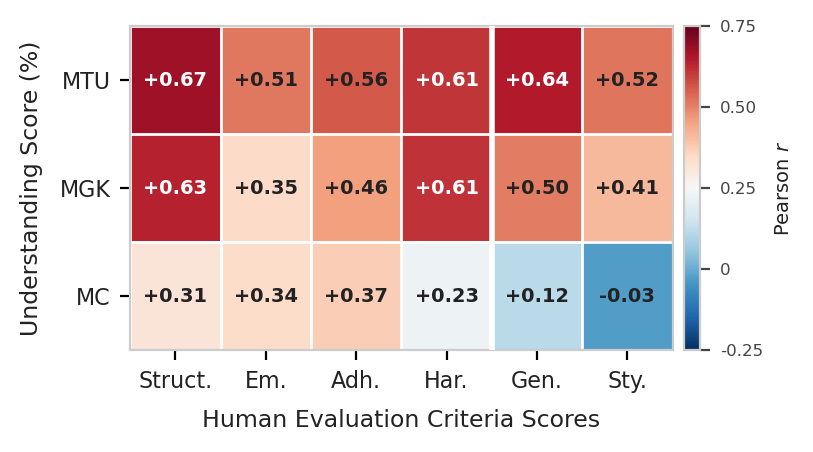}
\vspace{-3mm}
     \caption{Pearson correlation between Music understanding tasks vs Human evaluation criteria scores on the generation task for the top 9 models. MTU=Music Theory Understanding, MGK=Music General Knowledge, MC=Music Continuation; the x-axis presents structure, emotion, adherence to instruction, harmoniousness, genre, and style accuracy, respectively.}
    \label{fig:corr}
  \vspace{-5mm}  
\end{figure}

We further examined whether improved musical understanding translates into improved generation quality in the models.
Figure \ref{fig:corr} shows Music Theory Understanding and Music General Knowledge have moderate positive correlations with ordinal human evaluation criteria (Structureness, Emotion, Adherence to instruction and Harmoniousness) ($r = 0.35-0.67$), suggesting stronger theoretical knowledge tends to produce structurally sounder and emotionally richer outputs. However, correlations with Genre and Style Accuracy are weaker, and Music Continuation is essentially uncorrelated with Style Accuracy ($r = -0.03$), indicating that structural recognition ability does not predict generative stylistic competence.

\vspace{-1mm}
\subsubsection{Limitations in Automatic Evaluation}
\vspace{-1mm}
\cite{kader2025surveyevaluationmetricsmusic} identified persistent challenges including weak correlation between automatic metrics and human judgment, cross-cultural bias, and the absence of standardized evaluation frameworks in their survey of music generation evaluation. Similar concerns have emerged in other generation tasks. Benchmarking work on Semantic Overlap Summarization found substantial variation in how automatic metrics align with human judgments and emphasized the importance of correlation analysis when validating evaluation measures \cite{salvador-etal-2025-benchmarking}. Our findings provide empirical evidence for these concerns in the context of South Asian classical music, where existing symbolic metrics fail to capture stylistic faithfulness. To assess whether automatic metrics reliably capture stylistic faithfulness, we computed Pearson correlations between each automatic metric and human Style Accuracy across all evaluated models (Figure~\ref{fig:style}). Among five metrics, only KL Divergence shows a statistically significant correlation ($r = -0.73$, $p < 0.05$), yet even the strongest model (Gemini 2.5 Pro) reaches only 40\% style accuracy. Scale Adherence and ABC Syntax Accuracy show near-zero correlations ($r = -0.41$ and $r = -0.06$, both non-significant), models achieving 99-100\% Scale Adherence score as low as 0\% on Style Accuracy, confirming that syntactic correctness is entirely uncorrelated with stylistic faithfulness. Answering \textbf{RQ-4}, existing metrics operate at the syntactic and pitch-distribution level and can be satisfied trivially while producing outputs that annotators unanimously classify as stylistically unrecognizable. This mirrors a broader pattern in evaluation research: as with ALIGN-SIM \cite{mahajan-etal-2024-align}, strong performance on surface-level metrics does not reflect the underlying competencies those benchmarks intend to measure. Developing metrics sensitive to raga-specific ornamentation and the compositional conventions of Bengali classical traditions, and more broadly, culturally grounded evaluation frameworks that move beyond syntactic correctness, remains an open problem.

\vspace{-2mm}
\section{Conclusion}
\vspace{-1mm}
We present a systematic evaluation pipeline of 33 open-source and frontier LLMs on South Asian symbolic music understanding and generation, through a Bengali classical music benchmark spanning music theory, cultural knowledge, symbolic continuation, and ABC notation generation. Our results reveal
a pronounced capability gap: Gemini 2.5 Pro achieves \textbf{90.8\%},
\textbf{89.4\%}, and \textbf{85.2\%} across the three understanding subtasks,
while most
open-source models remain in the \textbf{23-40\%} range. However, we observed a critical gap between structural generation competence and fine-grained stylistic faithfulness:  despite reaching \textbf{95\%} South Asian genre adherence, produces recognizable Rabindra/Nazrul-style melodies only \textbf{40\%} of the time. Furthermore, existing automatic metrics operating at the syntactic and pitch-distribution level are insufficient for culturally grounded evaluation, highlighting the need for reliable evaluation of non-Western symbolic music as an open problem. These findings suggest that while frontier LLMs offer a promising foundation for low-resource South Asian music modeling, achieving faithful stylistic control and developing culturally grounded evaluation frameworks for music generation remain critical challenges for the field.

\section*{Limitations}
In this work, the generation evaluation focuses exclusively on two Bengali classical traditions, Rabindra and Nazrul Sangeet, which, while culturally significant and computationally underrepresented, do not fully represent the breadth of South Asian or even Bengali musical heritage. The scope of the Music Theory Understanding subtask is similarly constrained to Hindustani classical theory as practiced in the Bengali tradition. Thus, results on this benchmark should not be generalized to South Asian music competence broadly. Our choice of ABC notation, while motivated by both modeling and data constraints, imposes a ceiling on stylistic faithfulness that is independent of model capability. This information loss occurs at two levels: Swaralipi itself is a skeletal scaffold that hints at ornamentation through hyphens and slur marks but does not prescriptively encode \textit{gamakas}, \textit{meend}, or microtonal inflections, as these are transmitted through oral performance traditions. ABC notation further approximates these hints through grace notes and ties, but continuous pitch glides and microtonal inflections remain inexpressible. Richer formats such as MIDI or MusicXML could better capture such ornamentations. Unfortunately, direct conversion from Swaralipi into MIDI or MusicXML is infeasible because its solfège-based representation lacks the absolute pitch and timing encodings required by these formats. Consequently, reported style accuracy scores reflect faithfulness within ABC's expressive bounds, and the true gap between generated outputs and authentic raga performance practice may be larger than our evaluation captures.



\bibliography{references}

\appendix
\section{Benchmark Examples} \label{benchmark_examples}
Table \ref{tab:benchmark_examples} shows one example per sub-task from the South Asian Music Understanding Benchmark. 

\begin{table*}[!htb]
\centering
\scriptsize
\renewcommand{\arraystretch}{1.15}
\begin{tabular}{p{0.25\textwidth} p{0.71\textwidth}}
\hline
\textbf{Benchmark Sub-Task} & \textbf{Example} \\
\hline

Music General Knowledge &
\foreignlanguage{bengali}{ঠুংরী গানের সর্বপ্রথম প্রচলন হয় কোথায়?}\\
& A. \foreignlanguage{bengali}{বানারসী}\\
& B. \foreignlanguage{bengali}{আগ্রা}\\
& C. \foreignlanguage{bengali}{দিল্লী}\\
& D. \foreignlanguage{bengali}{লক্ষনৌ}\\
& What is the correct answer? Strictly output only the answer as option A or B or C or D in this following format: [[Answer:]]\\[0.5ex]
& \textbf{Translation:}\\
& Where was Thumri music first popularized?\\
& A. Benares\\
& B. Agra\\
& C. Delhi\\
& D. Lucknow\\
\hline

Music Theory Understanding &
\foreignlanguage{bengali}{কোনটি সত্য নয়?}\\
& A. \foreignlanguage{bengali}{ঠাট কখনো সাত স্বরের কম বা বেশি দিয়ে রচিত হয়না।}\\
& B. \foreignlanguage{bengali}{ঠাট কেবল অবরোহী হয়।}\\
& C. \foreignlanguage{bengali}{ঠাটের সংখ্যা মাত্র ১০ টি।}\\
& D. \foreignlanguage{bengali}{১০ টি ঠাট থেকেই সকল রাগের সৃষ্টি।}\\
& What is the correct answer? Strictly output only the answer as option A or B or C or D in this following format: [[Answer:]]\\[0.5ex]
& \textbf{Translation:}\\
& Which of the following is not true?\\
& A. A Thaat is never composed with fewer or more than seven notes.\\
& B. A Thaat is only descending in nature.\\
& C. There are only 10 Thaats.\\
& D. All ragas originate from these 10 Thaats.\\
\hline

Music Continuation &
Please select the most matching melody continuation segment based on the entered melody:\\
& X: 1\\
& T: \foreignlanguage{bengali}{হলুদ গাঁদার ফুল }\\
& M: 3/4\\
& K: C\\
& |: (D2 F-F) D2 | (F2 G) | A3 |\\
& | (D  F-F) D2 | (F2 G) | A3 |\\
& | A2 (A\_B | A) (G F) | (F G) (GA |G) (F E)|\\
& | (E F E) | (DC \_BA,2) | (A, C2-C) (C D) |\\[0.5ex]
& A. | B cB AB ABA | (GA G3) | GA G FE D | EG3 GG |\\
& | A B c ced | c3 | C C D EF | (FGF E3) |\\
& | F G \{G\}AG F | (\{F\}GF E2) DC | D EF \{D\}E DED | C4 |\\[0.5ex]
& B. | (E F) (E | D) D2 | \{D\}C3-C3 :|\\
& | G2 G-G G2 | A3 | G3 |\\
& | E (E G) | G G2 | A3 | G3 |\\
& | (G A) (A| c) (c d) | c\_B3 | A AG2 |\\[0.5ex]
& C. | (GA G) F | E F2-F2 D\_E | \_E (\_E D) |\\
& | z2 CD | D (C B,) |: C (\{B\}C D\_E) | \_E D2 |\\
& | D (D \_E) | F G2 | F (F \_E) | D (D\_E F\_E) |\\[0.5ex]
& D. | \_A2 \_A2 | B2 c c | \_d3 c | (cB -c) (\_B -\_A) :|\\
& | \_A \_e \_e \_e-\_e \_e (\_e \_d)| \_d \_d c2 | \_B \_B c2 |\\
& | (\_d3 -C) | c\_B c\_B (\_B \_A) |\\
\hline

\end{tabular}
\caption{Examples from each of the subtasks of the South Asian Music Understanding Benchmark used for evaluating Music General Knowledge, Music Theory Understanding, and Music continuation. Figure \ref{fig:subtheme} the theme-wise number distribution of questions of the benchmark.}
\label{tab:benchmark_examples}
\end{table*}

\begin{figure}[!htb]  
\centering
\includegraphics[width=\columnwidth]{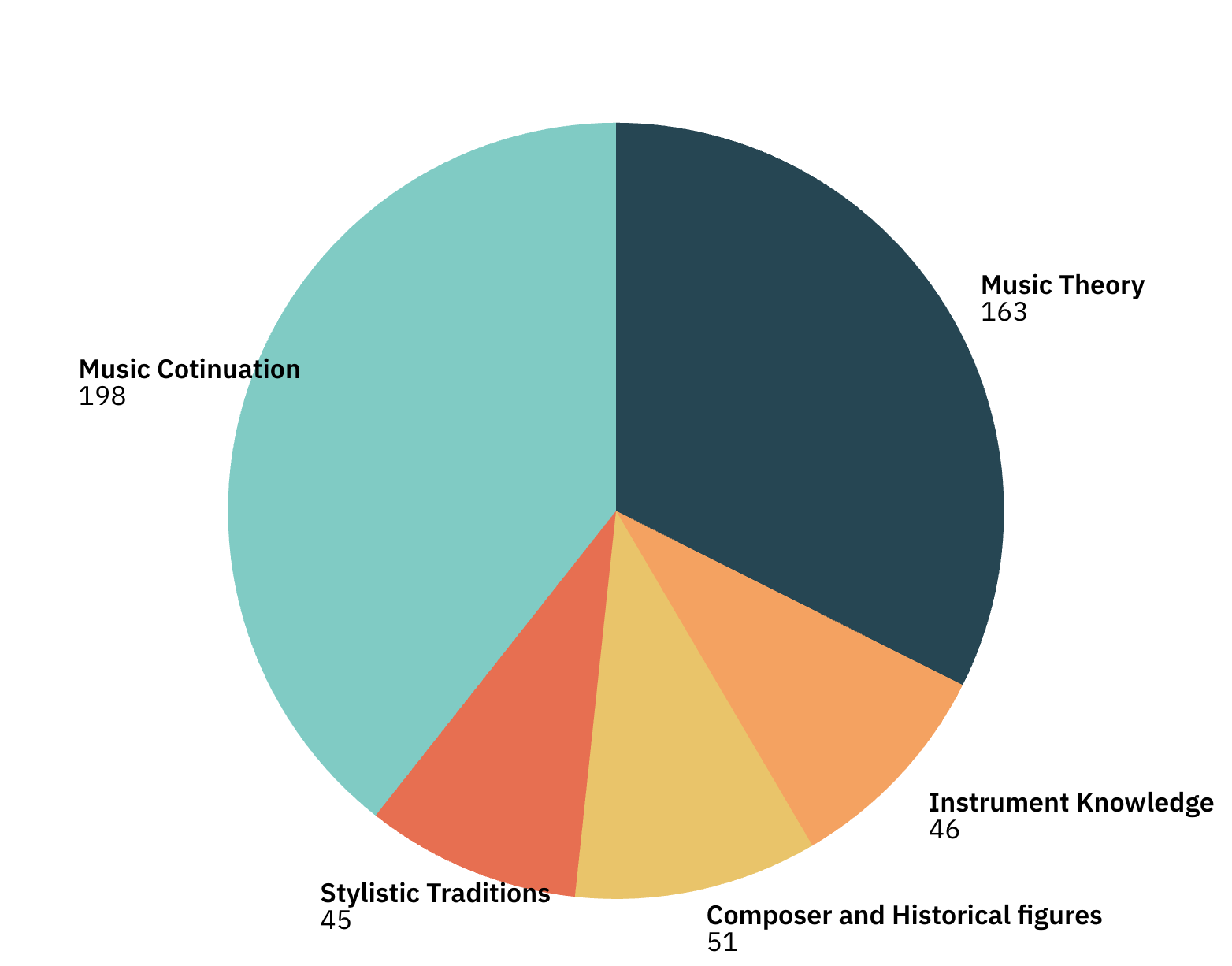}  
    
     \caption{Number of questions per sub-theme}
    \label{fig:subtheme}
    
\end{figure}

\section{Prompts for Music Generation} \label{prompt}
Table \ref{tab:abc-prompt-levels} shows the exact curated 5 level of prompts with increasing amount of details following the TELeR taxonomy \cite{santu2023telergeneraltaxonomyllm} for music generation in ABC notation. The meta-data that was provided were lyrics, theme, time signature from the original songs. Since the official Bengali Swaralipi notation system represents pitches using solfège 
syllables \foreignlanguage{bengali}{(সা, রে, গা, মা, পা, ধা, নি)} relative to a tonic without specifying an absolute 
pitch, a mapping convention was required during the manual transcription of Swaralipi scores 
into ABC notation. As ABC notation requires an explicit key signature declaration 
(\texttt{K:}), we adopted the convention of mapping \foreignlanguage{bengali}{সা} (Sa, the tonic) to C, yielding the 
correspondence: \foreignlanguage{bengali}{সা}=C, \foreignlanguage{bengali}{রে}=D, \foreignlanguage{bengali}{গা}=E, \foreignlanguage{bengali}{মা}=F, \foreignlanguage{bengali}{পা}=G, \foreignlanguage{bengali}{ধা}=A, \foreignlanguage{bengali}{নি}=B. All 100 ground 
truth ABC notations were transcribed under this fixed mapping, resulting in all scores being 
rooted in Scale C. Consequently, all music generation prompts were constrained to Scale C 
to ensure a consistent and fair basis for comparison between the generated outputs and the 
ground truth scores. 
\begin{table*}[!htb]
\centering
\scriptsize
\setlength{\tabcolsep}{6pt}
\renewcommand{\arraystretch}{1.15}
\begin{tabularx}{\textwidth}{p{0.18\textwidth} L}
\toprule
\textbf{Level} & \textbf{Prompt} \\
\midrule

\textbf{Level 1 -- Minimum Details} &
Generate a melody in ABC notation for the full lyrics [lyrics]. \\

\midrule
\textbf{Level 2 -- Moderate Details} &
Generate a melody in ABC notation for the full lyrics [lyrics], in Scale C with 3/4 time signature, without any comments and empty lines. The song will have a indian classical music theme and cover the theme of the lyrics which is about [theme] and feature instruments used in indian classical music. The song should be about 2--3 minutes length. \\

\midrule
\textbf{Level 3 -- Moderate Listwise Details} &
Generate a melody in ABC notation for the given lyrics by performing the following tasks:

\begin{enumerate}[leftmargin=*,itemsep=0.2ex,partopsep=0.2ex,parsep=0.2ex]
  \item Ensure the melody adheres to a Indian classical music style.
  \item It should sound like a [Rabindra/Nazrul] Song.
  \item Use Scale C and [time signature] time signature.
  \item Strictly generate just notations, avoid any comments and extra line.
  \item Cover the lyrical theme: [theme], ensuring the output spans \(\sim\)2--3 minutes in play length.
  \item The music should feature instruments used in indian classical music.
\end{enumerate}

[Lyrics] \\

\midrule
\textbf{Level 4 -- Significant Details + User Expectation} &
Level 3 prompt +\newline
A good output should:

\begin{itemize}[leftmargin=*,itemsep=0.2ex,partopsep=0.2ex,parsep=0.2ex]
  \item Be musically coherent and playable.
  \item Clearly reflect the theme [theme].
  \item Maintain stylistic authenticity of Indian classical music.
  \item Be structured in valid ABC notation syntax.
\end{itemize}

Lyrics:\newline
[lyrics] \\

\midrule
\textbf{Level 5 -- Maximum Details} &
Level 4 prompt +\newline
Additional background: Indian classical music typically involves raga-based melodies which works with a specific set of notes, ornamentations (gamakas), microtonal variations and an emotional or spiritual mood that matches the lyrics well and rhythmic cycles (taals). Unlike Western classical music, Indian classical tradition emphasizes monophonic texture -centered on melodic development rather than harmonic progression.\newline
Lyrics:\newline
[lyrics] \\

\bottomrule
\end{tabularx}
\caption{Prompt levels for ABC melody generation from lyrics.}
\label{tab:abc-prompt-levels}
\end{table*}

\begin{table}[!htb]
\centering
\footnotesize
\resizebox{\columnwidth}{!}{%
\begin{tabular}{lcccc}
\toprule
\textbf{Metric} & \textbf{L2} & \textbf{L3} & \textbf{L4} & \textbf{L5} \\
\midrule
KL Divergence $(\downarrow)$          & \textbf{3.911} & 4.320 & 4.232 & 3.971 \\
Repetition Rate (\%) $(\uparrow)$     & 0.047 & \textbf{0.196} & 0.107 & 0.080 \\
Scale Adherence (\%) $(\uparrow)$     & 0.901 & \textbf{0.923} & 0.860 & 0.884 \\
Pitch Histogram Entropy               & 2.477 & 2.300 & 2.394 & 2.411 \\
ABC Syntax Accuracy (\%) $(\uparrow)$ & 0.438 & \textbf{0.520} & 0.397 & 0.434 \\
\bottomrule
\end{tabular}}

\caption{Mean automatic evaluation scores across all 9 models per prompt level (L2--L5).}
\label{tab:mean_auto_eval}

\end{table}

\section{Answer Extractor} \label{answer}
To extract the one letter answer (A,B,C or D) for the multiple choice question, first, the model response is optionally stripped of the original prompt to avoid matches from the question or answer choices. The system then attempts to extract a definitive answer using a set of regular-expression patterns that capture explicit answer declarations (e.g., ``[[Answer: C]]”, ``The correct answer is B”, or ``Option D is correct”). If a match is found, the corresponding option (A-D) is returned and the prediction is deemed unambiguous. When no explicit pattern is detected, a fallback strategy is applied in which all standalone mentions of answer letters (A–D) are identified and counted. If a single option occurs most frequently, it is selected as the model’s prediction; if multiple options are tied for the highest frequency, the prediction is marked as ambiguous. Finally, if no valid answer option can be extracted from the output, the response is classified as ambiguous with no predicted answer. The ambiguous matches were checked manually afterwards for definitive answers.

\section{Automatic Evaluation Metrics} \label{auto}
\subsection{KL Divergence}
KL Divergence measures the statistical distance between two probability distributions, 
quantifying how much one distribution differs from a reference distribution, where a lower 
value indicates greater similarity between the two distributions. We computed the KL Divergence between the pitch distributions of the original and generated 
ABC notation scores as a reference-based evaluation metric. For each song, pitches were 
extracted from the ABC strings using a regular expression pattern that captures standard note 
names (A-G), their octave modifiers, and accidentals, followed by normalization to remove 
octave indicators and convert to uppercase. A pitch histogram $P$ was then constructed for 
each score by computing the relative frequency of each pitch class, formally defined as 
$P(x) = \frac{c(x)}{\sum_{x'} c(x')}$, where $c(x)$ denotes the count of pitch $x$. To 
measure how much the generated pitch distribution $Q$ diverges from the original distribution 
$P$, we computed the KL Divergence as:

\begin{equation}
\small
    D_{KL}(P \| Q) = \sum_{x} P(x) \log \frac{P(x)}{Q(x)}
\end{equation}

To handle zero-probability pitch classes, a small smoothing term $\epsilon = 10^{-10}$ was 
added to both distributions before normalization, ensuring numerical stability. The final 
reported score is the average KL Divergence across all song pairs within each model and 
prompt level, where lower values indicate that the generated pitch distribution more closely 
resembles the original.

\subsection{Pitch Histogram Entropy}
Pitch Histogram Entropy measures the diversity and uniformity of pitch usage in a melody, 
where a higher value indicates a more varied and evenly distributed use of pitches across 
the generated score.

We computed the Pitch Histogram Entropy of the generated ABC notation scores as a 
reference-free evaluation metric. For each generated score, pitches were extracted using 
the same regular expression pattern described above, normalized to remove octave indicators 
and converted to uppercase. A pitch probability distribution was then constructed by 
computing the relative frequency of each pitch class $x$ as $P(x) = \frac{c(x)}{\sum_{x'} 
c(x')}$, where $c(x)$ denotes the count of pitch $x$. The entropy of the pitch 
histogram was then computed as:

\begin{equation}
\small
    H = -\sum_{x} P(x) \log_2 P(x)
\end{equation}

The final reported score is the average entropy across all generated scores per model and 
prompt level. A higher entropy value reflects greater melodic diversity, suggesting the 
model utilizes a wider and more balanced range of pitches, whereas a lower entropy indicates 
a tendency toward repetitive or pitch-restricted melodic output.

\subsection{Repetition Rate}
Repetition Rate measures the proportion of generated scores that contain explicit repeat 
signs, where a higher value indicates greater use of structured musical repetition, which 
is a characteristic feature of well-formed compositions.

We computed the Repetition Rate of the generated ABC notation scores as a reference-free 
evaluation metric reflecting structural organization. For each generated score, we checked 
for the presence of the ABC notation repeat sign \texttt{|:}, which denotes a repeated 
section in the score. The Repetition Rate was then computed as:

\begin{equation}
\small
    \text{Repetition Rate} = \frac{\text{\# of scores containing } \texttt{|:}}{\text{Total scores}}
\end{equation}

The final reported score is the proportion of scores per model and prompt level that contained at least one repeat sign. A higher repetition rate suggests that the model generates more structurally coherent melodies with intentional repeated sections, which  aligns with the compositional conventions of both Rabindra and Nazrul Sangeet where returning melodic phrases such as the sthayi are a defining structural characteristic.

\subsection{ABC Syntax Accuracy}
ABC Parse Success Rate measures the proportion of generated scores that are syntactically 
valid and parseable as ABC notation, where a higher value indicates that the model produces 
more well-formed and structurally correct musical scores.

We computed the ABC Parse Success Rate of the generated scores as a reference-free 
evaluation metric reflecting syntactic correctness. For each generated ABC notation file, 
we attempted to parse and convert it to MIDI using the \textbf{music21} library\footnote{\url{https://music21.org/music21docs/}}. If the parse succeeded and a valid MIDI file was produced, the score was counted as a successful 
parse; if an exception was raised during conversion, the file was counted as a parse 
failure. The ABC Parse Success Rate was then computed as:

\begin{equation}
\small
    \text{ABC Parse Success Rate} = \frac{\text{\#of successfully parsed scores}}{\text{Total scores}}
\end{equation}

Successfully parsed files were subsequently converted to WAV audio using \textbf{FluidSynth}\footnote{\url{https://github.com/FluidSynth/fluidsynth}} 
with a SoundFont file, enabling both syntactic validation and downstream human evaluation 
of the generated audio. The final reported score is the proportion of scores per model and 
prompt level that were successfully parsed, where a higher rate indicates the model 
generates more syntactically valid ABC notation that conforms to the formal grammar 
expected by standard ABC parsers.

\subsection{Scale Adherence}
Scale Adherence measures the proportion of generated scores that strictly conform to the 
specified target scale, where a higher value indicates that the model generates melodies 
that remain within the prescribed set of allowable pitches.

We computed the Scale Adherence of the generated ABC notation scores as a reference-free 
evaluation metric reflecting pitch-level instruction following. Since all prompts specified 
Scale C, we defined the allowed pitch set as the C major scale $\{$C, D, E, F, G, A, B$\}$ 
along with a set of permitted flat pitches $\{$$\flat$E, $\flat$A, $\flat$B$\}$ 
that are characteristic of certain Indian classical ragas within that scale context. For 
each generated score, pitches were extracted and normalized using the same procedure 
described above. A score was considered scale-adherent if and only if every extracted pitch 
belonged to the allowed pitch set, with any occurrence of a sharp (\texttt{\^{}}) or 
natural sign (\texttt{=}) prefix immediately marking the score as non-adherent. The Scale 
Adherence rate was then computed as:

\begin{equation}
\small
    \text{Scale Adherence} = \frac{\text{\# of scores fully adhering to Scale C}}{\text{Total scores}}
\end{equation}

The final reported score is the proportion of scores per model and prompt level that passed 
the scale adherence check. A higher score indicates the model more reliably respects the 
scale constraint provided in the prompt, which is a necessary condition for stylistic and 
theoretical alignment with the target musical tradition.

\section{Annotation Guideline} \label{criteria}
Annotators were first briefed about the project and how the collected data was going to be utilized. Consent was taken on data usage and no personal information were collected during the evaluation. The annotators were given 9 anonymous folders including 20 samples each, not disclosing the models' names. The instruction given to the annotators is written verbatim below:

Each annotator receives three items: A folder named ``Audio files'', An Excel file (annotator\#.xlsx) and this instructions document.

\noindent \textbf{1. Folder structure:} Inside the ``Audio files'' folder, there are 9 subfolders. Each subfolder corresponds to a different large language model (LLM) and contains 20 audio files generated by that model. The names of these 9 subfolders exactly match the sheet names in the Excel file.

\noindent \textbf{2. Matching folders and sheets:} In annotator\#.xlsx, there are 9 sheets, each named after one of the subfolders. When you work on a specific folder, make sure you are also filling out the sheet with the same name. \textit{Example:} If you are annotating audios from the folder ``Folder 1'', fill in the sheet named ``Folder 1''.

\noindent \textbf{3. Understanding the Excel file:} Each row in a sheet represents one audio file. The ``song no.'' column specifies the exact file name you need to work on. The full file path is shown as: \textit{folder\_name/song\_name.wav}. \textit{Example:} ``Folder 1/5.wav'' means the file is inside ``Folder 1''.

\noindent \textbf{4. Annotation procedure:} Find the audio file mentioned in the ``song no.'' column. Read the prompt in the same row of the Excel sheet, this describes what the model was asked to generate. Listen carefully to the corresponding audio file. Fill in the remaining columns in that row based on the instructions given below.

\textbf{Note:} Even if the lyrics are well-known Najrul/Rabindra songs, as LLMs are asked to generate new melodies, the new melody might be totally different from the original song that we are used to listening to. Annotators should assess based on how the new melody aligns with the given theme and lyrics.

Each of the cells includes a drop-down menu to select your answer from. The drop-down button will show up on the top-right corner of the cell when you select it.

\noindent \textbf{Annotation Guideline:} Given a prompt including lyrics, theme, scale and time signature and their corresponding generated melody, listen to the melody and score it according to the criteria below.

\paragraph{Structureness}
Is the melody structured nicely and has distinct sections like:
\begin{itemize}
    \item \textbf{sthayi} (\foreignlanguage{bengali}{স্থায়ী}): Functions like the ``base'' section to which the singer returns after improvisations.
    \item \textbf{antara} (\foreignlanguage{bengali}{অন্তরা}): Expands the melody, usually in a higher octave; provides lyrical variation before returning to sthayi and can be paired with the sthayi (the performance alternates between them).
\end{itemize}

\begin{table}[h]
\centering
\footnotesize
\resizebox{\columnwidth}{!}{%
\begin{tabular}{lp{9cm}}
\toprule
\textbf{Score} & \textbf{Description} \\
\midrule
Unstructured &
    The melody is linear and repetitive, with no clear return point or section contrast. No identifiable sthayi or antara. \\
Slightly Structured &
    There are minor variations or pauses suggesting different ideas, but sections are not clearly distinguished or cyclic. \\
Moderately Structured &
    The melody shows partial organization -- a recognizable main idea (sthayi) with one contrasting section, though transitions may be unclear. \\
Well Structured &
    The melody clearly alternates between sthayi and antara, with evident octave or lyrical variation before returning to the base theme. \\
Highly Structured &
    The melody exhibits strong, distinct sections -- sthayi and antara are clearly defined, cyclic, balanced, and thematically coherent throughout. \\
\bottomrule
\end{tabular}}
\label{struct}
\end{table}

\paragraph{Genre Analysis}
Can the melody be classified as Indian classical music, including
characteristics such as microtones, ornamentations/gamakas, and
intricate tunes? Answer \textit{Yes} or \textit{No}.

\paragraph{Style Analysis}
Can the melody be classified as having a feel like Rabindra
Sangeet (\foreignlanguage{bengali}{রবীন্দ্র সঙ্গীত}) or Nazrul Sangeet
(\foreignlanguage{bengali}{নজরুল সঙ্গীত})? Answer \textit{Rabindra},
\textit{Nazrul}, or \textit{None}.

\paragraph{Emotion}
Does the emotion of the melody align with the given lyrics? 
\bigskip
\bigskip
\bigskip
\begin{table}[h]
\centering
\footnotesize
\resizebox{\columnwidth}{!}{%
\begin{tabular}{lp{9cm}}
\toprule
\textbf{Score} & \textbf{Description} \\
\midrule
Not Aligned at All &
    The melody conveys a completely different or conflicting emotion compared to the lyrics (e.g., cheerful tune for sorrowful lyrics). \\
Weakly Aligned &
    A few emotional cues match, but overall the melody feels disconnected or mismatched with the lyrical sentiment. \\
Moderately Aligned &
    The melody captures the general mood but lacks depth or consistency in expressing the intended emotion. \\
Well Aligned &
    The melody successfully supports and enhances the lyrical emotion, showing consistency in tone and phrasing. \\
Strongly Aligned &
    The melody deeply embodies the lyrical emotion -- dynamics, phrasing, and tonal quality fully reinforce the sentiment of the lyrics. \\
\bottomrule
\end{tabular}}
\label{emotion}
\end{table}
\bigskip
\bigskip
\bigskip
\paragraph{Adherence to Instruction (REL)}
How much does the melody adhere to the given instructions? 

\begin{table}[h]
\centering
\footnotesize
\resizebox{\columnwidth}{!}{%
\begin{tabular}{lp{9cm}}
\toprule
\textbf{Score} & \textbf{Description} \\
\midrule
Not at All &
    The melody ignores the provided instructions entirely -- wrong scale, rhythm, emotion, or structure; unrelated to the given prompt. \\
Slightly Adhering &
    Some aspects of the instructions are present (e.g., correct tempo or partial scale), but major elements like theme or mood are missing or inconsistent. \\
Moderately Adhering &
    The melody reflects several key aspects of the instructions (scale, emotion, or timing) but lacks precision or full consistency. \\
Well Adhering &
    The melody follows most of the given instructions accurately -- correct scale, fitting time signature, and thematic relevance with minor deviations. \\
Fully Adhering &
    The melody strictly and consistently follows all given instructions (scale, time signature, emotion, theme, structure) with high fidelity and musical coherence. \\
\bottomrule
\end{tabular}}
\label{rel}
\end{table}

\paragraph{Harmoniousness}
Is the melody properly in tune, harmonious, and interesting?.

\begin{table}[!htb]
\centering
\footnotesize
\resizebox{\columnwidth}{!}{%
\begin{tabular}{lp{9cm}}
\toprule
\textbf{Score} & \textbf{Description} \\
\midrule
Not Harmonious &
    The melody sounds off-key, dissonant, or unpleasant. Pitches clash frequently, creating instability or discomfort. \\
Weakly Harmonious &
    Some notes or intervals fit the scale, but tuning inconsistencies or awkward jumps make it sound rough or unbalanced. \\
Moderately Harmonious &
    The melody generally stays in tune and follows harmonic expectations, but lacks richness or engaging transitions. \\
Well Harmonious &
    The melody is tuneful and balanced; intervals and phrasing flow naturally with few discordant moments. Pleasant but not deeply striking. \\
Highly Harmonious \& Engaging &
    The melody is perfectly in tune, musically rich, and emotionally engaging. It maintains consonance, variation, and aesthetic appeal throughout. \\
\bottomrule
\end{tabular}}
\label{harm}
\end{table}
\end{document}